\documentclass[pre,twocolumn,showpacs]{revtex4}
\pacs{5.45.-a, 42.50.Vk}
\usepackage{graphicx}
\usepackage{amsmath}
\usepackage{amsfonts}
\usepackage{mathrsfs}
\newcommand{\kbar}{\mathchar'26\mkern-9muk}
\begin{document}
\title{Deviations from early--time quasilinear behaviour for the quantum kicked rotor near the classical
limit}
\author{Mark Sadgrove}
\author{Terry Mullins}
\author{Scott Parkins}
\author{Rainer Leonhardt}
\affiliation{Department of Physics, The University of Auckland, Private Bag 92019,
Auckland, New Zealand}
\begin{abstract}
We present experimental measurements of the mean energy for the
atom optics kicked rotor after just two kicks. The energy is
found to deviate from the quasi--linear value for 
small kicking periods. The observed deviation is 
explained by recent theoretical results which include the
effect of a non--uniform initial momentum distribution, previously
applied only to systems using much colder atoms than ours.
\end{abstract}

\maketitle

For some time now, studies of cold atoms subject to a periodically
pulsed 1D optical lattice (a system referred to as the atom
optics kicked rotor (AOKR)) have provided a great deal of insight into quantum
systems with chaotic classical analogues \cite{Moore1995, Ammann1998, Ringot2000, dArcy2001}.
Typically, such experiments have focussed on the ``late
time'' behaviour of the quantum kicked rotor. This is partially because the hallmark
of quantum interference effects in
the AOKR -- dynamical localisation -- is only observed after at least
five or ten kicks for typical experimental parameters (see for example \cite{Bharucha1999}).
Observations of quantum resonances in the mean energy have also generally been made in the late
time regime \cite{dArcy2001, Sadgrove2004}.
Somewhat less attention hasbeen given, however, to the ``early time'' regime, where theoretical
predictions have suggested that only quasi--linear energy growth takes place, in particular, 
during the first two kicks. That is, no classical or quantum correlations are expected to be important
during this time, given broad initial distributions for position and momentum.

However, recent theoretical and experimental work demonstrates that non-trivial 
behaviour may be observed in the mean energy of the AOKR as a function of 
kicking period after as few as two kicks if the atomic sample initially has
a narrow momentum distribution \cite{Daley2002}. Experimental work has focussed
on the temporal half Talbot effect which leads to recurrence of the initial momentum distribution
after two kicks \cite{Deng1999, Duffy2003} and more detailed studies of the energy dependence
as the kicking period is varied have also been performed \cite{Duffy2003}.
To observe the effects of interest, 
both of these previous experimental studies made use of Bose-Einstein condensates which provide
atomic samples with much narrower initial momentum distributions than those used in typical
AOKR experiments. For atomic ensembles with 
larger thermal energies, the classical theory of Rechester and White \cite{Rechester1980} and
the quantum theory of Shepelyansky \cite{Shepelyansky1987} both predict the same constant
quasi--linear energy growth rate for the first two kicks for any kicking period.
In this report we demonstrate that deviations from quasi--linear behaviour can occur in the second kick
even for relatively broad initial momentum distributions. The anomalous energy growth rates are found
only at very small values of the kicking period which have not previously been probed
experimentally.

For large detunings between the kicking laser and the atomic transition,
 the scaled Hamiltonian for an atom which experiences ideal $\delta$ kicks with period $T$ is
\begin{equation}
\hat{H}=\frac{\hat{\rho}^2}{2} - \kappa\cos(\hat{\phi})\sum_{n=0}^N\delta(\tau-n),
\label{eq:ham}
\end{equation}
where $\hat{\rho}$ is the scaled atomic momentum operator, $\hat{\phi}$ is the scaled position operator
 for an atom,
$\tau = t/T$ is the scaled time and $\kappa$ is the kicking strength. We note the commutator relation
$[\hat{\phi},\hat{\rho}] = i \kbar$ where $\kbar = 8\omega_r T$, and $\hbar \omega_r$ is the 
energy change of a Caesium atom after the scattering of a 
single photon of wavelength $2\pi/k_l = 852$nm has occurred. 
The atomic momentum $p$ is 
related to the scaled momentum $\rho$ by $p/2\hbar k_l = \rho/\kbar$,
and we refer to the quantity $p/2\hbar k_l$ as the momentum in 2-photon recoils, or
 the momentum in experimental units.
The atomic position operator $\hat{x}$ is given by $\hat{x} = \hat{\phi}/2k_l$.

It is useful to consider the standard map for the $\delta$--kicked rotor (DKR).
If we label the atomic position and momentum just before the $n$th kick as $\phi_{n-1}$ and
$\rho_{n-1}$ respectively, integrating Hamilton's equations over one kick gives the recursive 
relation:
\begin{subequations}
\begin{eqnarray}
\phi_n & = & \phi_{n-1} + \rho_n \\
\rho_n & = & \rho_{n-1} - \kappa\sin(\phi_{n-1}).
\end{eqnarray}
\label{eq:stdmap}
\end{subequations}
This map holds for the classical position and momentum for the DKR and also
for the associated position and momentum operators for the quantum DKR.
In this report, we are interested in the
mean energy of the atoms after they have experienced two kicks. From Eq. \ref{eq:stdmap}
 the atomic momentum after two kicks is
\begin{equation}
\rho_2 = \rho_0 - \kappa\sin(\phi_0) - \kappa\sin(\phi_1).
\label{eq:mom2kicks}
\end{equation}
Experimentally, we measure the quantity $E_2 = \langle\rho_2^2/2\kbar^2\rangle$, that is, the mean
kinetic energy 
of the atomic ensemble (in experimental units). Theoretically, this requires the determination of 
correlation
functions of the form $\langle \sin(\phi_0)\sin(\phi_i) \rangle$. These correlations were
first calculated for the classical DKR by Rechester and White \cite{Rechester1980} and 
later for the quantum DKR by Shepelyansky \cite{Shepelyansky1987} under the assumption
that the atomic position and momentum were initially uniformly distributed.
In this case, the cross terms in the expression for
\begin{figure}[tbh]
\centering
\includegraphics[height=6.5cm]{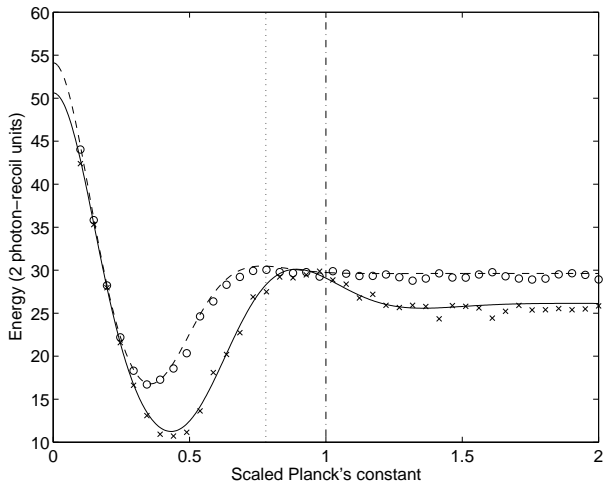}
\caption{Energy after 2 kicks as a function of $\kbar$ as given by Eq. \ref{eq:daley2kicks}.
 The solid line is for $\kappa/\kbar = 7$ and $\sigma_p=1.8$ 2--photon recoils.
The dashed line is for $\kappa/\kbar = 7$ and $\sigma_p=3.2$ 2--photon recoils.
Simulation results for the same parameters (crosses and circles respectively), but for kicking with 
rectangular pulses of constant width $\tau_p$, are overlayed. The ratio $\tau_p/T$ 
varied from about 0.015 for $\kbar\approx 3$ to 0.5 for $\kbar \approx 0.1$. Vertical
lines show the value of $\kbar_{\rm crit}$ as determined from Eq. \ref{eq:deviate}. 
The dot-dashed line
is for $\sigma_p=1.8$, and the dotted line is for $\sigma_p=3.2$.
 \label{fig:sim_theory}}
\end{figure}
$\langle\rho_2^2\rangle$ are found to vanish when averaged over the ensemble and we are left only
 with contributions from the squares of the three terms in Eq. \ref{eq:mom2kicks}. The last
two terms give equal contributions of $\kappa^2/4$ each, so the energy
after two kicks is
\begin{equation}
E_{2,{\rm broad}} = \frac{1}{2\kbar^2}(\sigma_\rho^2 + \kappa^2),
\label{eq:Ebroad}
\end{equation}
where $\sigma_\rho$ is the scaled standard deviation of the initial momentum distribution.
Although this expression is explicitly dependent on $\kbar$, in a given experimental run, the 
quantities $\sigma_\rho/\kbar$ and $\kappa/\kbar$ are held constant 
(corresponding to a constant MOT temperature and constant laser power respectively).
In this case, we see that the energy after two kicks remains constant as $\kbar$ is varied 
and that, ignoring the thermal energy
 $\sigma_\rho^2/2\kbar^2$, it is simply given by 
twice the quasi-linear growth rate $\kappa^2/4\kbar^2$. 
\begin{figure}[tbh]
\centering
\includegraphics[height=6.5cm]{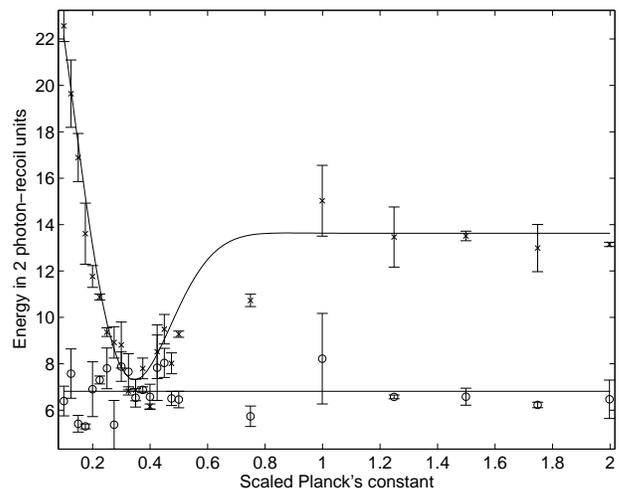}
\caption{Experimentally measured energies after one and two kicks (circles and crosses 
respectively). The value of $\kappa/\kbar$ was found to be $5.2 \pm 0.4$ and $\sigma_p$
was measured to be 4.2 2-photon recoils. Solid lines
give the theoretical values for the energy after one and two kicks respectively.
The initial thermal energy has been subtracted to allow a more instructive comparison between the 
energies.  \label{fig:onetwo}}
\end{figure}

The assumption of 
a broad initial momentum distribution is not justified if the initial momentum
distribution is comparable in width to $2\hbar k_l$ (as is the case for a Bose-Einstein condensate),
\emph{or} if $\kbar$ is close to zero, since $\sigma_\rho = \kbar \sigma_p$ (where $\sigma_p$ is the 
standard deviation in experimental units). In this report we consider the latter case and the effect
 it has on
 the energy after two kicks.
In Ref. \cite{Daley2002} an expression was derived for the energy after the second kick for atoms
 with a
Gaussian initial momentum distribution of arbitrary width $\sigma_\rho$.
In this case the cross terms in the expression for $\langle\rho_2^2 \rangle$ do not vanish and the
 energy
after two kicks for the quantum DKR is found to be 
\begin{eqnarray}
E_2 & = & \frac{1}{2\kbar^2}\left(\sigma_\rho^2 + \frac{\kappa^2}{2} + 
\frac{\kappa^2}{2}(1 - J_2(\kappa_{2q})e^{-2\sigma_\rho^2}) \right. \label{eq:daley2kicks} \\
& & -\kappa J_1(\kappa_q)\sigma_\rho^2e^{-\sigma_\rho^2/2} + \kappa^2(J_0(\kappa_q) \nonumber \\
& & \left. - J_2(\kappa_q))\cos(\kbar/2)e^{-\sigma_\rho^2/2}) \right),  \nonumber
\end{eqnarray}
where $\kappa_q = 2\kappa\sin(\kbar/2)/\kbar$ and $\kappa_{2q} = 2\kappa\sin(\kbar)/\kbar$.
The width of the initial momentum distribution becomes narrower in scaled units 
as $\kbar$ is decreased, leading to deviations from the quasilinear
result for $\kbar \lesssim 1$. This effect can be seen in Fig. \ref{fig:sim_theory} for two
different values of $\sigma_p$. The energy now varies as a function of $\kbar$ and exhibits 
a pronounced minimum, the position of which depends on the exact value of $\kappa/\kbar$. 
From this minimum, the energy then
increases monotonically as $\kbar \rightarrow 0$ to the value $(\kappa/\kbar)^2 + \sigma_p^2/2$
in 2--photon recoil energies.

Physically, we can ascribe the importance of the exact momentum variance
of the atoms for low $\kbar$ to the short time between pulses in this regime. Essentially, the spread
in atomic momentum is not resolved if the time between the first and second kicks is small. For very 
small $\kbar$, the 
system's behaviour is similar to the case where the atomic sample starts in an initial 
momentum eigenstate. We may estimate the value of $\kbar$ below which we expect deviations from
quasi--linear behaviour to occur as follows: After one kick, the atoms have a momentum variance
$\sigma^2_{\rm tot} = \sigma_p^2 + \sigma_1^2$, where $\sigma_1$, the momentum variance due to the 
first kick,
is $\sigma_1^2 = 2(\kappa/\kbar)^2(\hbar k_l)^2$ (assuming a broad initial position distribution).
The shortest time $T_{\rm crit}$ between pulses for which atoms with momentum 
$|p| = \sigma_{\rm tot}$ still traverse a full cycle of
the standing wave between kicks is given by 
\begin{equation}
T_{\rm crit} = \frac{\lambda M_{\rm Cs}}{2\sigma_{\rm tot}}.
\label{eq:deviate}
\end{equation}
For pulsing periods $T < T_{\rm crit}$ 
(and, thus, $\kbar<8\omega_rT_{\rm crit}=\kbar_{\rm crit}$), the majority of the atoms
traverse a distance less than $\lambda/2$ and deviations from the quasilinear result
should be expected. For the
two sets of parameters considered in Fig. \ref{fig:sim_theory}, $\kbar_{\rm crit}=1.0$
and $0.78$, in good agreement with the behaviour exhibited by numerical
and analytical results.

The crosses and circles in Fig. \ref{fig:sim_theory} show simulation results for a
rectangular-pulse-kicked rotor, in which the pulse width $\tau_p$ was kept constant as the kicking period
was decreased, as in experiments. We see that even though the $\delta$--kicked approximation is flagrantly 
violated for low $\kbar$ (where the pulse may be on for up to half the kicking period),
the simulation results show excellent agreement with the $\delta$--kicked theory. This 
agreement may be attributed to the fact that the atomic momenta are still sufficiently small after
 two 
kicks, that atoms will not tend to traverse a significant distance along the optical standing wave
during the time the pulse is on. Specifically, for the highest energies measured in this work,
atoms typically travel a distance of only $8\%$ of the standing wave's period. This means that 
averaging of the kicking strength over the pulse duration, which can restrict energy growth after
 many pulses,
may be neglected in this work. Hence, we
compare our experimental results with Eq. \ref{eq:daley2kicks} in the remainder of this report.
\begin{figure}[tbh]
\centering
\includegraphics[height=6.5cm]{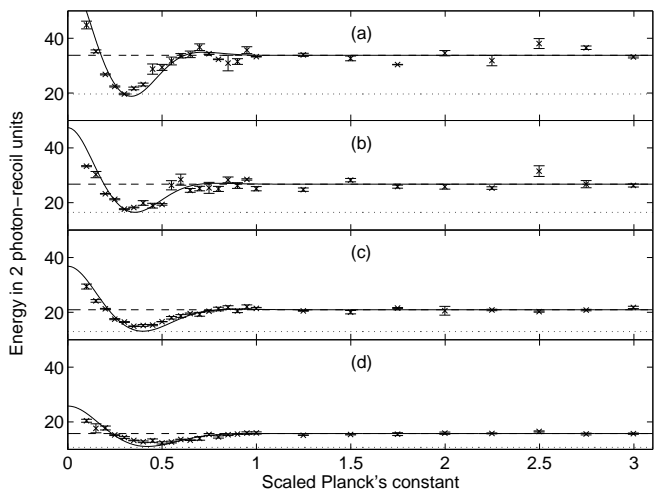}
\caption{Experimentally measured energies after two kicks for various values of
$\kappa/\kbar$ and $\sigma_p=3.35 \pm 0.06$.
 The measured values of $\kappa/\kbar$ were (a) 7.5, (b) 6.4, (c) 5.6 and (d) 4.5
each with an experimental error of $\pm 0.1$. The solid line through the data shows the 
energy given by Eq. \ref{eq:daley2kicks}. The dotted and dashed lines show the energy given
by the quasi--linear theory after one and two kicks respectively. \label{fig:power_change}}
\end{figure}

Our experimental setup has been detailed in other publications \cite{Williams2004, Sadgrove2004}.
Caesium atoms are typically cooled to below $10\mu$K for our experiments in a standard MOT
 \cite{Monroe1990}.
The atoms are then
released from the trap and kicked twice by pulses from an optical standing wave detuned
500 MHz from the $6S_{1/2}(F=4) \rightarrow 6P_{3/2}(F'=5)$ transition of Caesium. 
The pulse width $\tau_p$ remained constant at $480$ns for all experimental runs. After
a 12 ms expansion time, the atoms are effectively frozen in space by optical molasses and their
resultant fluorescence is captured by a CCD camera giving information about the position distribution
of the atoms after kicking. Knowing the time of flight of the atoms, we may infer 
their momentum distribution from the CCD image. We then find the energy of the atomic ensemble
by numerically calculating the second moment of this distribution..

Fig. \ref{fig:onetwo} shows experimental results after one and two kicks. For the results
shown in this figure, $\kappa/\kbar$ was measured by finding the mean of the energies 
when $\kbar \geq 1$ for both one and two kicks. The difference between these two values
is given by the quasilinear energy growth rate $1/4(\kappa/\kbar)^2$.
Using this parameter, and subtracting the thermal energy from both sets of results,
we find good agreement between the analytical formula of Eq. \ref{eq:daley2kicks} and
 the results obtained
experimentally.
Since the mean energy after one kick is expected to always be the trivial quasilinear result,
the remainder of our results
focus on the energy after two kicks. In this case, we measure $\sigma_p$ using a time of flight method
and once again find the mean energy for the results where $\kbar \geq 1$.
We may then calculate $\kappa/\kbar$ using Eq. \ref{eq:Ebroad}.
In Fig. \ref{fig:power_change}, the mean energy as a function 
of $\kbar$ is shown for various values of the parameter $\kappa/\kbar$ corresponding to different
laser powers used in our experiment. We notice in particular that the position in $\kbar$ of the
energy minimum and the point where deviation from the quasilinear result first occurs shift
to the right as the value of $\kappa/\kbar$ is decreased.
Additionally, it is interesting to note the value that the analytical expression tends to
as $\kbar \rightarrow 0$, even though this region was not able to be probed experimentally.
In this limit, the energy due to kicking tends to $(\kappa/\kbar)^2$, which corresponds
to ballistic energy growth with coefficient $1/4(\kappa/\kbar)^2$. 
Ballistic energy growth 
is usually associated with quantum resonance and, indeed, the $\kbar \rightarrow 0$ limit 
may be seen as a special case of quantum resonance. In fact, the $\epsilon$--classical picture
developed for the usual quantum resonances of the kicked rotor \cite{Wimberger2003} should
also be applicable in the regime near $\kbar = 0$. We anticipate investigating the energy peak
as $\kbar \rightarrow 0$ in more detail in the near future.
\begin{figure}[tbh]
\centering
\includegraphics[height=6.5cm]{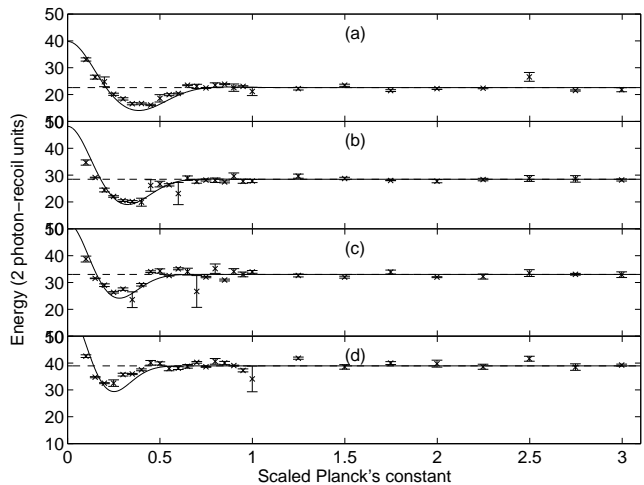}
\caption{Experimentally measured energies after two kicks for various values of
$\sigma_p$. The measured values of $\sigma_p$ were (a) 3.3, (b) 4.2, (c) 5.3 and (d) 6.0
in 2-photon recoils.
We estimate the error in these measurements to be in the order of $0.9$ 2-photon recoil.
For these curves, the measured value of $\kappa/\kbar$ is $5.8 \pm 0.2$.
Solid lines show analytical results and dashed lines show the energy predicted by the
quasi--linear theory.
\label{fig:tempchange}}
\end{figure}

We have also investigated the effect of increasing $\sigma_p$ on the deviation from quasi--linear 
behaviour. We would expect that as $\sigma_p$ gets larger, the deviation from the quasi--linear 
energy in the second kick would become less prominent. To test this experimentally,
we reduced the cooling efficiency of our MOT to create atomic samples with various momentum spreads
of up to $6.0$ 2--photon recoils. These samples were then kicked for the same value of $\kappa/\kbar$.
As shown in Fig. \ref{fig:tempchange} the deviations do indeed become less pronounced 
as the initial momentum spread is increased. However, modern kicked rotor experiments
typically achieve initial momentum distributions much narrower than 6 2-photon recoils, so
experiments studying the early--time behaviour for small $\kbar$ clearly need to take
finite--$\sigma_p$ effects into account.

\begin{figure}[tbh]
\centering
\includegraphics[height=6.5cm]{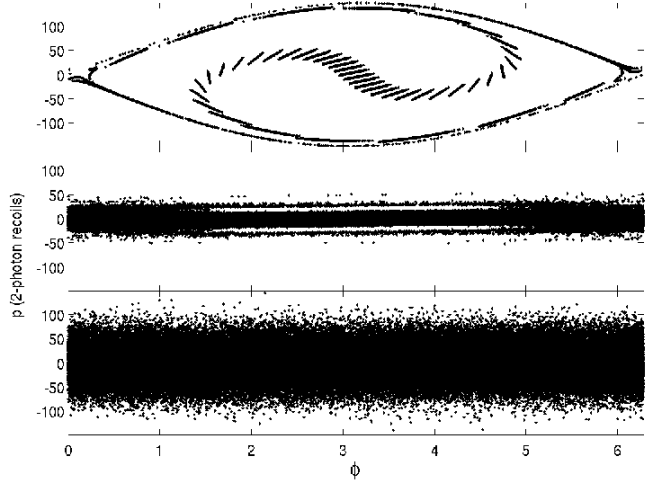}
\caption{Phase space diagrams for $T$ corresponding to $\kbar =$ (top) $0.001$, (middle) $0.3$ and
(bottom) $3.0$ and $\kappa$ such that $\kappa/\kbar=5.5$. The plots were generated from 100 
iterations of the map in Eq. \ref{eq:stdmap}
using a uniform initial distribution in the $\phi$ coordinate and a Gaussian
initial distribution in $p$ (momentum in experimental units) with $\sigma_p = 3.6$. \label{fig:phasespace}}
\end{figure}

The structure seen in our results occurs for $\kbar \lesssim 1$, where it seems reasonable to
expect that the classical kicked rotor theory is of considerable relevance.
In Fig. \ref{fig:phasespace}, the classical phase space distribution is shown
for kicking periods corresponding to $\kbar = 0.001$, $0.3$  and $3.0$ and 
values of $\kappa$ such that $\kappa/\kbar = 5.5$.
 The trend in the momentum spread (and thus energy) as $\kbar$
increases is apparent and does indeed appear to match that found in our measurements, 
in particular, the dip in energy seen near $\kbar = 0.3$.
One factor governing this behaviour is the
transition from regular motion to chaos as $\kappa$ is increased in the kicked rotor system.
The onset of global chaos occurs for $\kappa \sim 1$ \cite{Lichtenberg1992}, corresponding
to $\kbar=0.18$ for the parameters used in Fig. \ref{fig:phasespace}. Thus the energy minimum 
would seem to correspond roughly to the transition to chaos where resonant motion has been destroyed 
but diffusive energy growth is still inhibited by Komolgorov--Arnold--Moser boundaries.
We note that the phase space diagram was obtained using 100 iterations of the standard map 
in order to emphasise the structure present. However, the qualitative behaviour of the system 
is seen to be the same as that after just two kicks.

In conclusion, we have presented new experimental results for the Atom Optics Kicked Rotor at 
very early times 
and as $\kbar \rightarrow 0$. Although, the $\delta$--kicked  approximation is no longer a good 
one for
experiments in this regime, we nonetheless find excellent agreement between our measurements and
 recent
analytical predictions.
These results deviate significantly from the quasi--linear energy growth in the first two kicks
predicted by theories which assume a broad initial momentum distribution.
The considerations raised in this report will be important for any future studies of the atom optics
kicked rotor in the regime of small $\kbar$. Additionally, the further study of this system for small 
$\kbar$ is of interest as it relates to quantum resonance behaviour in the kicked rotor.

The authors would like to thank Sandro Wimberger for enlightening discussions about this work.
M.S. is supported by a Top Achiever Doctoral Scholarship 03131. This work was supported by the Royal 
Society of New Zealand Marsden Fund, grant UOA016.

\end{document}